\documentclass[useAMS,usenatbib,usegraphix]{mn2e}
\usepackage{graphicx}
\usepackage{natbib}
\usepackage{color}

\title[On the origin of radio-loudness in AGNs]
{On the origin of radio-loudness in AGNs and its relationship with the 
properties of the central supermassive black hole}
\author[Marco Chiaberge and Alessandro Marconi]{Marco Chiaberge$^{1,2}$\thanks{E-mail:
marcoc@stsci.edu} and Alessandro Marconi$^{3}$\\
$^{1}$Space Telescope Science Institute, Baltimore, MD 21218\\
$^{2}$INAF - IRA, Via P. Gobetti 101, Bologna, I-40129\\
$^{3}$Dipartimento di Fisica e Astronomia, 
Universit\'a di Firenze, L.go E. Fermi 2, Firenze I-50125}
\begin{document}

\date{ }

\pagerange{\pageref{firstpage}--\pageref{lastpage}} \pubyear{2010}

\maketitle

\label{firstpage}

\begin{abstract}
We  investigate   the  relationship   between  the  mass   of  central
supermassive  black holes  (SMBH)  and the  radio  loudness of  active
galactic  nuclei.   We use  the  most  recent  calibrations to  derive
``virial'' black hole masses for  samples of radio loud QSOs for which
relatively small masses ($M_{BH}  < 10^8 M_\odot$) have been estimated
in  the literature.   We take  into  account the  effect of  radiation
pressure  on the  BLR  which reduces  the ``effective''  gravitational
potential experienced  by the broad-line  clouds and affects  the mass
estimates of bright quasars.  We  show that in well defined samples of
nearby  low luminosity  AGNs (LLAGN),  QSOs  and AGNs  from the  SDSS,
radio-loud AGN  invariably host  SMBHs exceeding $\sim  10^8 M_\odot$.
On the other hand, radio--quiet AGNs are associated with a much larger
range of black  hole masses. The overall  result still holds even
without correcting the BH mass  estimates for the effects of radiation
pressure. We present a conjecture  based on these results, which aims
at explaining the origin of radio-loudness in terms of two fundamental
parameters: the  spin of the black  hole and the black  hole mass.  We
speculate  that in  order  to produce  a  radio-loud AGN  both of  the
following  requirements must  be  satisfied: 1)  the  black hole  mass
M$_{BH}$ has to be larger than $\sim 10^8 M_\odot$; 2) the spin of the
BH must be significant,  in order to satisfy theoretical requirements.
Taking  into  account the  most  recent  observations,  we envisage  a
scenario  in which  the  merger history  of  the host  galaxy plays  a
fundamental role in accounting for  both the properties of the AGN and
the galaxy morphology,  which in our picture are  strictly linked.  On
the one hand, radio loud  sources might be obtained only through major
``dry''mergers involving  BH of large  mass, which would give  rise to
both  the ``core''  morphology  and the  significant  black hole  spin
needed. On the  other hand, radio quiet AGNs  might reside in galaxies
that underwent different evolutionary  paths, depending on their black
hole mass.

\end{abstract}


\section{Introduction}

Radio-loud (RL) and  radio quiet (RQ) AGNs exist  at all luminosities.
The distinction between  the two classes is usually  made by using the
so-called radio  loudness parameter $R$,  i.e.  the ratio  between the
radio flux at 5GHz and the optical flux in the B band.  While powerful
RL and RQ quasars quasars  typically separate at values of $R\sim 10$,
at the  lowest luminosities,  the transition occurs  at a  much higher
value  \citep[e.g][]{xu99,terashima03,papllagn,sikora07}.  The general
interpretation   is  that   the   output  of   radio-loud  nuclei   is
energetically dominated by the jet,  while that of radio-quiet AGNs is
mostly dominated by the accretion disc.  However, the physical reasons
for the origin of the  observed differences still remain unknown.  How
can an AGN develop a radio jet on either large scales (hundreds of kpc
or even larger)  or small (pc or sub-pc) scales?   Why some AGN posses
such energetically dominant jets and  others do not?  How are the mass
and the spin of the central SMBH related to the radio-loudness?  These
questions are central not only for a complete understanding of the AGN
phenomenon but  also to  assess the role  of the  central supermassive
black  hole in  the evolution  of the  galaxy, the  rise of  a ``radio
phase'' and  its impact on the  evolution of the galaxy  itself and on
the environment \citep[e.g.][]{best05,bower06,croton06}.

Jets are most  likely formed by extracting rotational  energy from the
black   hole  and   the   accretion  disc   through  magnetic   forces
\citep{blandfordznajek77,blandfordpayne82}.   Jet production  may also
help to  remove angular momentum  and drag material towards  the black
hole.  Understanding  under which physical  conditions accreting black
holes are capable of producing  some sort of collimated outflow is not
the  only critical  point.  It  is  more crucial  to understand  which
mechanism may be able to  produce a (relativistic) jet whose radiative
output  is a  significant fraction  of the  accretion  luminosity, and
which   may   even   become   the   dominant   source   of   radiation
\citep[e.g.][]{allen06,celottighisellini08}.  The jet power is somehow
related to  the spin of the  black hole, as the  B-Z mechanism clearly
establishes \citep{blandfordznajek77}.  However, other parameters such
as the magnetic  field and, possibly, the mass of  the black hole, may
also  come  into  play.    Significant  effort  has  been  devoted  to
investigate  such an  issue, but  no  general consensus  has yet  been
reached. \citet{franceschini98} found a tight correlation of the black
hole mass  with radio power in  a sample of  local AGNs; \cite{laor00}
studied a sample of PG QSOs, and set a limit for radio loud objects at
M$\sim   10^9\,M_\odot$,    a   similar   result    being   found   by
\citet{dunlop03}  and \citet{best05}.   \citet{xu99} pointed  out that
the distribution of L$_{[OIII]}$ for  radio loud AGN extends to higher
luminosities than that  of radio quiet sources, and  noted that such a
result  is consistent  with a  higher ``maximum''  black hole  mass in
radio loud AGNs.   However, other authors have found  evidence for the
opposite,  i.e.  there is  a significant  fraction of  radio-loud AGNs
associated    with   black    holes   of    relatively    small   mass
\citep[e.g.][]{ho02,woo02a,woo02b,rafter09}.

Clearly, our understanding of  the link between the radio-loudness and
the  BH  mass  critically depends  on  the  accuracy  of the  BH  mass
estimate.   Direct  BH  mass  estimates based  on  spatially  resolved
stellar and  gas kinematics  are possible only  in the  local universe
($D<\sim  200\,  Mpc$), and  their  complexity  does  not allow  their
application  to  large  samples  \citep[e.g.][]{ferrareseford05}. Estimates
of BH  masses in large  samples of objects  at all redshifts  are only
possible in AGNs with broad emission lines: BH masses are estimated by
applying  the  virial  theorem  $M_{BH}  =  f  \Delta  V^2  R_{BLR}/G$
\citep[see       e.g.][for      recent       reviews       on      the
subject]{peterson10,vestergaard10} where $f$  is a calibration factor,
$\Delta V$  is the broad line  width and $R_{BLR}$ is  the average BLR
size, usually  estimated from  the AGN continuum  luminosity following
the    $R_{BLR}-L$    relation    by    \citet[][see also \citealt{bentz09}]{kaspi00}.  It  is currently  believed that  the  accuracy of
these estimates is of the order of 0.3-0.5 dex r.m.s. \citep{peterson10}.
Recently, \citet{marconi08,marconi09} pointed out that BLR clouds are
subjected  to  radiation  pressure  from the  absorption  of  ionizing
photons and provided a simple  additive correction to the above virial
relation,  which is  proportional to  the continuum  luminosity.  Such
correction  increases  BH  mass  estimates  in  AGN  with  significant
luminosities compared to their BH mass.

In this paper, we build on the results shown in \cite{papllagn} and we
further investigate  the relationship between the mass  of the central
supermassive black  hole (SMBH) and  the radio loudness of  the active
nucleus  using  samples of  AGNs  at  all  luminosities, and  BH  mass
estimates obtained  with different methods.  In Sect.~\ref{samples} we
briefly   describe    the   samples    of   AGN   we    consider;   in
Sect.~\ref{estimates} we  describe the methods we use  to estimate the
mass of the central black hole; in Sect.~\ref{results} we describe the
results  and in  Sect.~\ref{discussion}  we discuss  our findings,  we
propose a possible scenario to interpret the results of this work, and
we draw conclusions.

\section{The samples}
\label{samples}
It is very important to investigate whether the mass of the central BH
plays a role  in determining the radio-loudness of  the associated AGN
in objects of all luminosities, from nearby nuclei with faint activity
to the most powerful quasars.  However, it is also extremely important
to discuss  objects of  different AGN powers  separately, in  order to
avoid misinterpreting the results. In the following sections we briefly
describe the AGN samples used in this paper.

\subsection{Low luminosity AGNs}

We consider the following samples of nearby low luminosity AGN:

{\bf 1)} The  complete sample of FR~I Radio Galaxies  at redshift $z <
0.1$  (i.e.   low luminosity  radio  galaxies)  from  the 3CR  catalog
\citep{3cr,pap1}. The 3C sample is selected in the radio band at a low
frequency (178MHz), therefore it is free from any orientation biases.

{\bf 2)} Seyfert~1 galaxies from the optically selected Palomar Survey
of nearby galaxies and from the CfA sample \citep{hopeng}.  We include
only  the Type~1 objects,  since the  line-of sight  to the  nuclei is
thought  to be obstructed  by dust  in those  belonging to  the Type~2
class.

{\bf 3)}  A complete, distance  limited ($d<19$ Mpc) sample  of LINERs
taken from the Palomar Survey of nearby galaxies \citep{ho97}.

{\bf 4)} 51  nearby early-type galaxies (E+S0) with  radio emission $>
1$ mJy at 5 GHz  (optical $+$ radio selection) \citep[][and references
therein]{sandrobarbara05}. The  large majority of the  galaxies in the
sample are spectroscopically classified as either LINER or Seyfert.  A
detailed    description     of    this    sample     is    given    in
\cite{sandrobarbara05}.

{\bf 5)} The 12 broad-line radio galaxies  with $z<0.3$
included in the 3CR catalog \citep[and references therein]{pap4}.

Samples 1, 2 and 3 have  been studied in  detail in \cite{papllagn}
and more details about those samples  can be found in that paper.  The
sample of nearby ellipticals (4) partially overlaps with samples 1, 2,
and 3.   However, there are only  10 objects in common,  so the total
number of  objects considered here  is 142.  Note that  being selected
according  to different criteria,  these objects  do not  constitute a
complete sample.  However, they  well represent the overall properties
of all kinds of low power active nuclei in the local universe.

We do  not discuss  in detail the  sample of \citet{ho02},  which also
claimed to find radio-loud AGNs  associated with low mass black holes,
for  two reasons.   Firstly, the  sample partially  overlaps  with the
\citet{hopeng}  and  \citet{ho97}  samples,  which we  consider  here.
Secondly, the  work by \citet{ho02}  includes a significant  number of
Type~2  AGNs. For all  sources, the  nuclear luminosity  was estimated
using an indirect  indicator (i.e. the luminosity of  the Hbeta line).
That might  produce results that are inconsistent  with those obtained
for the samples we include in this paper.

\subsection{QSO samples}

We  focus on  samples of  radio  selected radio-loud  QSOs taken  from
\citet{oshlack02} and \citet[][and references therein]{gu}, which were
found  to include  a significant  number of  SMBH with  estimated mass
lower  than  $\sim  10^8  M_\odot$.  The  sample  of  \citet{oshlack02}
comprises  flat-spectrum radio  loud quasars.  These objects  might be
significantly  affected by  relativistic beaming,  which  enhances the
radiation  both  in the  radio  and in  the  optical.   
The  redshifts of  all QSOs  in the  above samples  are in  the range
$0<z<1$.

We  also consider  the sample  of high-z  ($2.0 <  z <  2.5$)  QSOs of
\citet{mcintosh},    which   includes    very   luminous    ($L   \sim
10^{46}-10^{47}$  erg   s$^{-1}$)  quasars  of   both  radio-loud  and
radio-quiet class.

Note that  all of the  above samples of  quasars were included  in the
study of SMBH masses in AGN made by \citet{woo02a} and \citet{woo02b}.

\subsection{SDSS AGNs}

We also  include in  our analysis the  sample of ``broad  lined'' AGNs
from  \citet{rafter09}(and references therein), which  consists of  objects selected  from the
Sloan  Digital Sky  survey \citep[DR5,][]{sdssdr5}  with $z<0.35$,  for
which  radio counterparts  have been  found  in the  VLA FIRST  survey
\citep{first}.   The  sample  includes  a significant  number  of  low
luminosity  AGNs  ($L_{H\alpha}  <   10^{42}$  erg  s$^{-1}$)  and  is
extracted  from the  list originally  selected  by \citet{greeneho07}.
However, higher luminosity  objects ($L_{H\alpha} \sim 10^{43-44}$ erg
s$^{-1}$) are also represented in the sample.

\section{Methods for black hole \emph{mass} estimates}
\label{estimates}

The SMBH  masses for the  objects belonging to the  samples considered
here  are estimated using  different methods,  from gas  kinematics to
single epoch estimates based on scaling relations \citep[see e.g.][]{vestergaardrev09}.

The BH masses of LLAGNs  taken from \citet{papllagn} are derived using
either   the   relation   with   the   stellar   velocity   dispersion
\citet{tremaine02}  or   more  direct  measurements  from,  e.g.,   gas
kinematics taken from the literature.  For a fraction of the Seyfert~1
galaxies  (the  brightest  objects  belonging  to  that  sample),  the
estimates were made using reverberation mapping.  More details for the
samples  of FRIs,  Seyferts  and  LINERs can  be  found in  \citet[][and
references therein]{papllagn}.   
For the  samples of early-type galaxies  of \cite{sandrobarbara05} the
BH mass  estimates are made using the  relation of \citet{tremaine02}.
Black hole masses for 3CR BLRG are also derived using the same method,
with  the  only exception  of  3C390.3 for  which  we  used data  from
reverberation mapping \citep{kaspi00}.  For these LLAGN samples we use
the  SMBH  masses  from  the  literature  since  the  updated  scaling
relations  do not  provide  significantly different  values for  those
objects.

Furthermore, we note that the updated relation between the BH mass and
the  central velocity dispersion  provided by  \citet{gultekin09} does
not return values significantly different from those obtained with the
\citet{tremaine02} formula, for the purpose of this work.

In the following we will also make use of the relation between BH mass
and near-IR host  spheroid luminosity \citep{marconi03}. Such relation
has  a scatter  similar  to  the relation  with  the stellar  velocity
dispersion,     and    has     therefore     a    similar     accuracy
\citep[e.g.,][]{marconi03,graham07,hu09}.

The BH virial masses for all of the above QSO samples published in the
literature were derived using  single epoch estimates based on scaling
relations.  The  formulae typically use  the FWHM of a  broad emission
line  (usually H$\alpha$  or H$\beta$,  and more  rarely CIV)  and the
luminosity  of the  adjacent continuum  as crucial  parameters.  Those
relations are calibrated using calibrated using low-z Type~1 AGNs of a
broad  range of luminosities  for which  the BH  masses are  know from
reverberation      mapping      techniques      \citep[see][for      a
review]{vestergaardrev09}.

Here  we first  estimate the  BH masses  for all  QSOs using  the most
updated  formulae  \citep{vestpeterson06},  calibrated  to  up-to-date
reverberation    mapping   masses.     Furthermore,   as    noted   by
\citet{marconi08},  the  radiation  from  the  accretion  disc  exerts
pressure  on the  broad line  clouds which  opposes  the gravitational
attraction of the  black hole.  Therefore, in particular  for the more
luminous  quasars,   the  BLR   may  actually  experience   a  smaller
``effective'' gravitational  field. Thus the  mass of the BH  might be
underestimated when using single epoch estimates and the FWHM of broad
emission lines  such as H$\alpha$ and H$\beta$  as crucial parameters.
\citet{marconi08,marconi09}   calibrated  the  effects   of  radiation
pressure on the BLR. In this  paper we use the most up-to-date version
of  the radiation  pressure-corrected virial  relation to  estimate BH
masses (Marconi et  al.~2011, in preparation).  The formula  we use is
the following:

\begin{eqnarray}
\frac{M_{BH}}{M_{\odot}} &= & 10^{6.6}  \left(\frac{\rm{FHWM}(\rm{H}\beta)}{1000 \rm{km~s}^{-1}}\right)^2
\left(\frac{\lambda L_{\lambda}(5100\rm{\AA})}{10^{44} \rm{erg~s}^{-1}}\right)^{0.5} +\nonumber\\
 & & +10^{7.5} \times \left(\frac{\lambda L_{\lambda}(5100\rm{\AA})}{10^{44} \rm{erg~s}^{-1}}\right) 
\end{eqnarray}
where  we  use  H$\alpha$   when  measurements  of  H$\beta$  are  not
available. The  coefficients $10^{6.6}$ and  $10^{7.5}$, correspond to
the $f$  and $g$ coefficients in  \citet{marconi08}, respectively. The
values used  in this work  have been improved  based on new  data, but
they do not significantly differ from the original values.

The  presence  of  outward  radiation  forces  on  BLR  clouds  is  an
unavoidable physical effect due to  the injection of momentum from the
absorption of ionizing photons; its effect on virial mass estimates is
negligible only  if one makes  the (unlikely) assumption that  all BLR
clouds     have     very      large     column     densities     $N_H>
10^{24}\,\mathrm{cm}^{-2}$    \citep{netzer09,marconi09}.    Recently,
\cite{netzer10} studied  the motions of BLR clouds  under the combined
effects of  gravity and radiation pressure. They  concluded that, even
if radiation pressure is important,  BH masses derived from the simple
virial  product are not  significantly underestimated.   However, that
conclusion  strongly depends  on the  assumption that  BLR  clouds are
moving  in  pressure  equilibrium  within a  confining  medium,  whose
assumed pressure  gradient tunes  cloud column densities  during their
orbits.  All  clouds must  survive several dynamical  timescales (i.e.
must complete  several orbits) so  that on average the  virial product
will not  be affected  by radiation pressure.   Some kind  of magnetic
confinement is  invoked for the  clouds \citep[e.g.][]{ferland88}, but
overall  it is  not clear  from this  model how  can the  clouds avoid
Rayleigh-Taylor        and        Kelvin-Helmholtz       instabilities
\citep[e.g.][]{mathews87}.  Moreover  the only direct  observations of
the structure of BLR clouds based on eclipsing of the X-ray AGN source
suggest a cometary like structure, as expected from supersonic motions
of dense clouds in a less dense medium, and indicate a short lifetimes
of BLR clouds, less than  the orbital time scale \citep{maiolino10}. A
detailed discussion of  these issues will be subject  of a forthcoming
paper   (Marconi    et   al.~2011,   in    preparation).    See   also
Sect.~\ref{onmarconi} for a short  discussion about the impact of such
a correction on our results.

\section{Results}
\label{results}

\begin{figure}
\includegraphics[width=84mm]{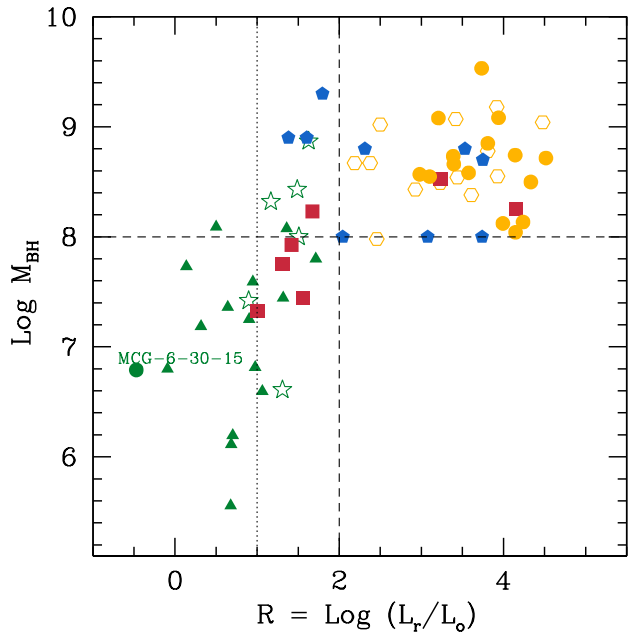}
\caption{Black  hole   masses  plotted  against   the  radio  loudness
parameter  R for  the sample  of  low luminosity  AGNs. Yellow  filled
circles are FRIs, green filled triangles are Seyferts, red squares are
LINERs, yellow  open hexagons are  core-galaxies and green  open stars
are  power-law galaxies  (see text).   The dotted line  shows the
``standard'' RQ/RL division based on PG quasars \citep{kellerman89}.}
\label{llagnfig}
\end{figure}

\subsection{Low luminosity AGNs}

In Fig.~\ref{llagnfig}  we plot the  estimated BH mass vs.   the radio
loudness  parameter  $R$  for   the  samples  of  nearby  LLAGN.   The
properties of the  objects belonging to that sample  are well defined,
thus  allowing us to  derive robust  results. Most  importantly, their
nuclear emission  can be  resolved by  using the VLA  and HST  for the
radio and  the optical bands,  respectively.  This allows us  a direct
comparison  between their  faint  nuclei and  the  nuclei of  powerful
quasars,  minimizing  the contributions  of  the  host galaxy  stellar
emission \citep[see][]{papllagn}.  Seyferts  are plotted as triangles,
power-law  galaxies  as stars,  LINERs  as  squares,  FR~Is as  filled
circles, core-galaxies\footnote{Core galaxies have luminosity profiles
that  rise steeply  towards  the  center, then  flatten  at a  certain
``break radius''. Power-law galaxies, instead, have profiles that rise
steeply all  the way to  the HST resolution  $\sim 0.1^{\prime\prime}$
citep{lauer95}.}  as empty circles, and BLRGs as pentagons.

The first  piece of information that  is important to bear  in mind is
that at  low AGN powers, the  ``separation'' between RQ  and RL nuclei
occurs  at  a  much  higher  value of  the  radio  loudness  parameter
\citep{papllagn,sikora07} than for powerful QSOs.  The reason for such
a  behavior is  still  unclear. \citet{sikora07}  showed  that such  a
separation  is a  function of  the  Eddington ratio  $L_o /  L_{Edd}$.
Therefore,  it  is  possible  that   a  change  in  the  nuclear  SED,
corresponding to, e.g.,   a change  in some of the physical properties of
the accretion disc,  might result in a different value  of $R$ for the
transition between a jet-dominated  and a disk-dominated AGN. However,
a  detailed analysis of  this subject  is beyond  the subject  of this
paper. 

The  dashed line  at $R=2$  in Fig.~\ref{llagnfig}  is drawn  with the
purpose of visually separating  objects for which the optical emission
is  disc-dominated  (radio-quiet, left-hand  side  of  the plot)  from
objects  that are jet-dominated  (radio-loud, right-hand  side).  Note
that BLRG  (blue pentagons)  are present in  both sides of  the plane.
This is  due to the  fact that BLRG  are objects seen  at intermediate
viewing  angles   with  respect   to  the  jet   direction  \citep[see
e.g.][]{barthel89,grandi07}.    Therefore,  most  likely   because  of
relativistic  beaming  effects,  in  some  of those  objects  the  jet
dominates the optical  emission, while in others the  jet radiation is
``de-beamed''  and the  disc is  bright enough  to overshine  the jet.
Their  location  may  also   be  affected  by  variability.   However,
independently  of   their  location,  it  is  clear   that  those  are
``intrinsically'' radio-loud  objects, i.e.  they  do produce powerful
relativistic  jets, irrespective  of the  observed  dominant radiation
source.

As already noted by \cite{papllagn},  all RL LLAGN are associated with
BH masses $>\sim 10^8$ M$_\odot$,  while most of the RQ population has
lower  BH  masses.  A  similar  result  has  been  recently  found  by
\citet{baldicapetti10}.  There seems to  be a ``region of avoidance'',
in the bottom-right  part of the plot, as  radio-loud LLAGN with small
black    hole    mass     are    absent\footnote{For    clarity,    in
Fig.~\ref{llagnfig}  we plot  only the  detected nuclei.  A  number of
objects with  upper limits to  the optical emission are  present among
the core galaxies and the LINER sample.  However, all core galaxies in
the Balmaverde sample  have BH mass $>\sim 10^8$,  therefore the exact
value of R  is irrelevant for the purpose of this  work. Only 3 LINERs
with detected radio emission  have undetected optical nuclei.  That is
most likely  explained as  due to the  high surface brightness  of the
central region  of the host  galaxies \citep[see e.g.   the discussion
in][]{capettib2}.  This leaves  our conclusions unaltered.}.  While it
is clear that radio quiet AGN  exist at all BH masses, the question is
whether radio loud AGN associated with small BH masses exist at all.

Note the location of the  Seyfert galaxy MCG--6--30--15 on the left hand
side  of   the  plane,  at  a   BH  mass  of  $\sim   6  \times  10^6$
\citep{mchardy05}.  This  is an  extremely important object,  since it is
probably the most compelling example of maximally spinning SMBH in AGN
\citep[e.g][]{iwasawa96,miniutti07}.   A high BH  spin has  been often
claimed to  be the origin  of radio-loudness \citep{blandfordsaasfee},
but MCG--6--30--15 is radio quiet. However, the BH mass is at least 1 dex
smaller than any RL AGN in these samples. This points to the idea that
the spin  alone cannot give rise  to radio loudness  and is consistent
with    the    suggestion     presented    in    this    paper    (see
Sect. \ref{discussion}).

\subsection{QSOs}

\begin{figure}
\includegraphics[width=84mm]{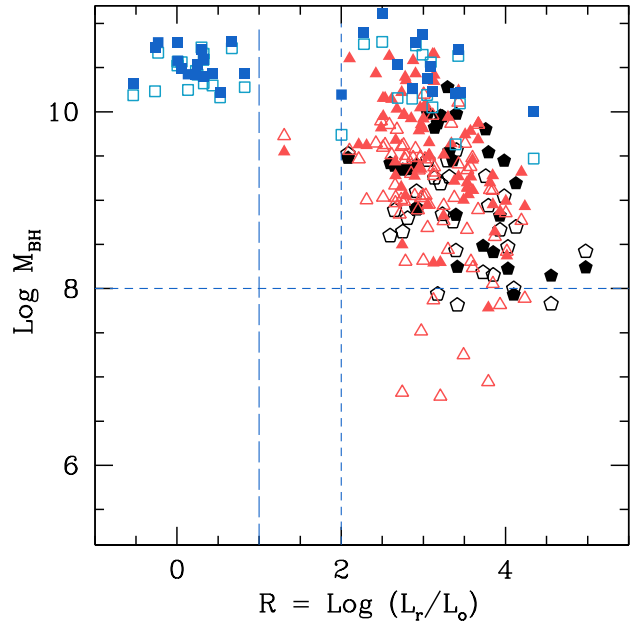}
\caption{Black hole  masses of QSO  samples plotted against  the radio
loudness parameter R. Empty symbols  refer to BH masses estimated with
the  \citet{vestpeterson06}  formula.   Filled symbols  are  radiation
pressure corrected masses  from \citet{marconi08}. Triangles, pentagons and
squares  represent  QSOs from  the  Gu et  al.,  Oshlack  et al.   and
McIntosh et al.  samples, respectively.}
\label{qsofig}
\end{figure}

While  it  is clear  that  the above  results  hold  for the  selected
(although well defined) sample of  objects, the question is whether it
can be  extended to larger samples  of objects and for  AGNs of higher
luminosity.   In order  to do  so,  in Fig.~\ref{qsofig}  we plot  the
samples  of QSOs  described in Sec.~2.2.   These are  particularly important
samples,  since relatively  small  black hole  virial masses  ($< \sim
10^8$  M$_\odot$)   have  been  estimated  in  the   literature  for  a
significant fraction of objects.

Firstly,  we  re-calculated  the  BH  masses for  all  QSOs  in  those
samples\footnote{We have  removed one objects  that was misclassified,
i.e.  the ``double  system'' PKS0114+074  \citep{akujor92}, and  a few
more for which the data in the literature are incomplete or unreliable
(none  of  those  are  associated  to BH  masses  smaller  than  $10^8$
M$_\odot$).} using the most updated formulae for virial mass estimates
derived  by \citet{vestpeterson06},  and adopting  the  WMAP cosmology
\citep{hinshaw09} $H_0 =71$  Km s$^{-1}$ Mpc$^{-1}$, $\Omega_M= 0.27$,
$\Omega_{vac}=  0.73$.   The results  are  shown  as  open symbols  in
Fig.~\ref{qsofig} (triangles,  pentagons and squares  represent the Gu
et al., Oshlack  et al.  and McIntosh et  al.  samples, respectively).
Note that  although the high-z  sample of \citet{mcintosh}  is clearly
biased against low BH masses, it  is useful to have it included in our
analysis in order  to show that RQ QSOs are  associated with both high
and low  BH masses and no physical  correlation between radio-loudness
and  BH  mass  exists,  as  already pointed  out  by  various  authors
\citet[e.g][]{woo02b}.

Although  the number  of objects  with  M$_{BH} <  10^8$ M$_\odot$  is
significantly smaller than found in the above cited papers as a result
of the updated formulae we used in this paper, a few objects are still
present in the  radio-loud and small BH mass region  of the plane (see
Fig.~\ref{qsofig}).   \citet{marconi08}   and  \citet{marconi09}  have
recently pointed  out that  for high luminosity  QSOs, the  effects of
radiation pressure  on the  broad line clouds  (in particular  for the
Hydrogen  lines) should  be  included  in the  BH  mass estimate.   We
applied  the  most  updated  corrections  (Marconi et  al.   2011,  in
preparation)   and we  calculated the  BH  mass for  all objects.   The
results are plotted as filled symbols in Fig.~\ref{qsofig}.  It is now
clear that all  QSOs in these samples, both RQ and  RL, have BH masses
above a certain threshold, somewhere close to $10^8$ M$_\odot$.

\begin{figure}
\includegraphics[width=84mm]{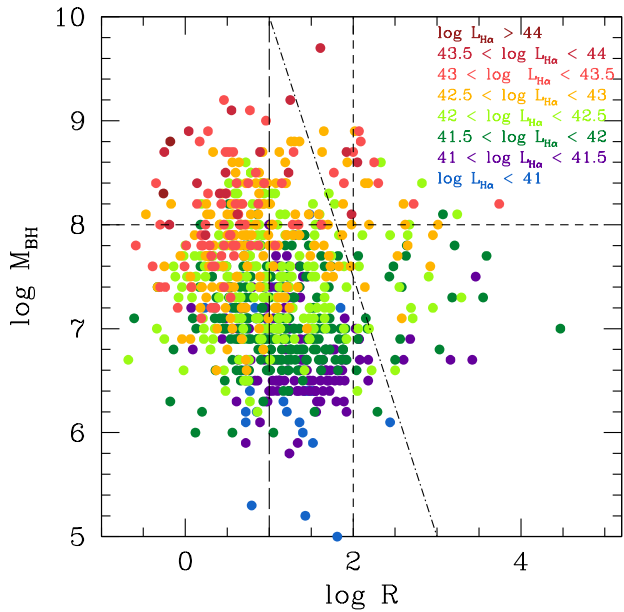}
\caption{Black  hole mass  vs radio  loudness parameter  for  the SDSS
sample  of  \citet{rafter09}.  The  color  coding  refers  to bins  of
H$\alpha$ luminosity,  which is used as  a rough indicator  of the AGN
power. The long dash line indicates the \citet{kellerman89} separation
between RQ  and RL QSOs. The  diagonal dot-dash line is  used to guide
the eye and follow the change in RQ/RL division with AGN luminosity.}
\label{rafterfig}
\end{figure}

\subsection{SDSS AGNs}

We  also want  to  check that  larger  samples of  AGNs selected  with
respect  to  their  optical  spectroscopic properties  behave  as  the
samples of AGNs  discussed above.  In order to do  so, we consider the
sample  from the  SDSS  published in  \citet{rafter09}.   A number  of
``radio-loud'' AGNs associated with small BH masses have been found in
that  sample.  The  BH  masses were  estimated  using the  H${\alpha}$
 FWHM, under the assumption that  the optical continuum
flux at 5100\AA~  is indicative of the AGN  luminosity. Note that
this assumption may not be  true for the lowest luminosity objects, in
which  the stellar  emission from  the  host galaxy  may dominate  the
optical flux at that wavelength.  

As already pointed  out above, at low luminosities  and for low values
of the Eddington ratio, the RQ/RL  division occurs at or above $\log R
\sim 2$  \citep{papllagn,sikora07}, therefore the number  of bona fide
``radio loud'' AGN in the Rafter et al.  sample should be reconsidered
taking  into account  the luminosity  class of  the objects.   In Fig.
\ref{rafterfig}  we plot  the BH  mass versus  radio loudness  for the
sub-sample of AGN detected in the radio band from \citet{rafter09}. We
use  a color  coding for  different luminosity  classes, and  we adopt
adopt the luminosity of the  broad component of the H$\alpha$ emission
line as a  (rough) indicator of the AGN power.  First  of all, we note
that the most luminous AGNs (L$_{H\alpha} >\sim 3 \times 10^{43}$ erg
s$^{-1}$) are all  associated with large BH masses  M$_{BH} >\sim 10^8$
M$_\odot$.  The objects appear to  be clustered in the RQ region (left
hand side) and the peak of the AGN radio-loudness distribution appears
to  shift  from  $\log$ R$\sim  2$  for  AGNs  of very  low  H$\alpha$
luminosities,  to  $\log$  R$\sim  1$  at  the  highest  luminosities.
Assuming  log R  = 1  as  the RQ/RL  threshold, the  percentage of  RL
objects  increases from $\sim  20\%$ for  objects with  $L_{H\alpha} >
10^{43}$  erg  s$^{-1}$  to  65\%  in the  luminosity  bin  between  $
L_{H\alpha} =  10^{41}$ erg s$^{-1}$  and $L_{H\alpha} =  10^{42}$ erg
s$^{-1}$.   Even  if  we  cannot  infer  the  actual  fraction  of
radio-loud AGNs  from that  sample, it is  unlikely that  the apparent
lack  of  high  luminosity  radio  loud AGNs  (M$_{BH}  >  8$,  R$>1$,
L$_{H\alpha} > 10^{42}$ erg s$^{-1}$) with respect to lower luminosity
objects is  due to a selection  effect.  In fact, the  sample could in
principle be biased against distant radio quiet objects, but the radio
loud ones  are certainly not affected.  Therefore,  that just confirms
that the  RQ/RL dividing  line shifts towards  higher values of  R for
objects of decreasing luminosity.  In fact,  if we assume $\log$ R = 2
for the RQ/RL division in the lowest luminosity bin, the percentage of
RL  objects returns to  a more  reasonable value  of $\sim  10\%$. The
diagonal dot-dashed line in  Fig.~\ref{rafterfig} is only used for the
purpose of guiding the eye and  follow the change in the RQ/RL divide.
However,  it is clear  that the  limit at  R=2 we  used for  the LLAGN
samples described  above well represents  the RQ/RL division  for this
sample, except possibly for the brightest and the faintest objects.

Nevertheless, a significant number of objects with $\log R > 2$ seem to
be associated with relatively  small black hole masses.  Therefore, we
select the objects that have R$> 2$ and M$_{BH}$ significantly smaller
than  $10^8$ M$_\odot$  (because of  the  uncertainty in  the BH  mass
estimates, we set our limit at $10^{7.5}$ M$_\odot$).  We find 36 such
objects, and we carefully  check their overall properties. The results
of  the analysis  is discussed  in  the following,  and summarized  in
Table~\ref{tabnewmasses}.

\begin{table*}
 \centering
 \begin{minipage}{140mm}
  \caption{Data for AGNs  with R$>$2 and $\log$ M$_{BH}  < 7.5$ in the
  SDSS sample  of \citet{rafter09}. }
  \begin{tabular}{lcccccr}
  \hline
  SDSS Name &  z  & R & $\log$ M$_{BH}$ & u-r  & $\log$ M$_{BH}$ & Notes \\
 \hline
 J003443.51-000226.6     &    0.042  &    2.44    &       6.1    &    2.31   &     8.3$^c$ &   \\
 J075244.19+455657.3     &   0.0517  &    2.07    &       6.9    &    2.73   &     8.8$^c$ &   \\
 J075444.08+354712.8     &    0.257  &    2.55    &       7.4    &    1.09   &     7.8$^d$ &   \\
 J083045.21+370946.7     &    0.155  &    2.15    &       7.1    &     2.3   &     8.9$^c$ &   \\
 J085010.42+074758.5     &     0.18  &    3.55    &       7.3    &    2.43   &     8.9$^c$ &   \\
 J090307.84+021152.2     &    0.329  &    2.14    &       7.4    &    2.44   &       --    &  QSO2, ULIRG  \\
 J090615.53+463619.0     &   0.0847  &    3.16    &       6.7    &    2.87   &     8.8$^c$ &   \\
 J092936.73+571149.8     &    0.262  &    2.04    &       6.6    &    1.74   &        --     & Type 2  \\
 J093712.33+500852.1     &    0.276  &    3.29    &       7.3    &    1.23   &     8.29$^a$ &   \\
 J094003.77+510421.8     &    0.207  &    2.56    &       6.9    &     2.9   &     9.0$^c$ &   \\
 J094525.90+352103.6     &    0.208  &    3.18    &       7.3    &    1.41   &       --      & Type 2/Galaxy?  \\
 J100410.85+523025.1     &    0.299  &    2.06    &         7    &     1.4   &     8.76$^a$ &   \\
 J103143.51+522535.1     &    0.167  &    2.69    &       7.2    &    2.56   &     8.8$^c$ &   \\
 J103330.65+070407.3     &    0.141  &     2.6    &       6.7    &    2.25   &     8.2$^b$ &   \\
 J103915.69-003916.9     &    0.077  &    2.18    &       6.7    &    2.14   &       --      &  Type 2/Galaxy?  \\
 J104029.16+105318.2     &    0.136  &    2.03    &         7    &    2.33   &     8.9$^c$ &   \\
 J110845.48+020240.8     &    0.158  &    2.71    &       7.1    &     2.3   &     9.0$^c$ &   \\
 J111807.47+002734.9     &    0.169  &    2.08    &       6.7    &    2.87   &     8.87$^a$ &   \\
 J115409.27+023815.0     &    0.211  &    2.58    &       7.4    &    2.48   &     7.7$^d$ &   \\
 J115437.43+114858.9     &     0.33  &    2.95    &       7.2    &    0.48   &    8.01$^a$ &   \\
 J122209.29+581421.5     &   0.0998  &    2.68    &       6.9    &    2.71   &     8.5$^c$ &   \\
 J124651.26+150914.3     &    0.323  &    3.03    &       6.8    &    3.22   &     8.9$^b$ &   \\
 J124707.32+490017.8     &    0.207  &    4.47    &         7    &    2.51   &     9.1$^c$ &   \\
 J130633.04+002248.4     &    0.148  &    2.07    &       6.9    &    2.43   &     8.8$^c$ &   \\
 J135646.10+102609.0     &    0.123  &    2.22    &       6.6    &    1.71   &       --    & Type 2    \\
 J140638.22+010254.6     &    0.236  &    2.17    &         7    &    1.32   &       --  & Outflow   \\
 J142237.91+044848.5     &    0.087  &    3.75    &       3.9    &    2.07   &       --    & Type 2 FWHM(H$\alpha$)=700    \\
 J144341.53+383521.8     &    0.162  &    2.29    &       6.2    &    1.94   &       --    & Type 2    \\
 J151513.58+552504.2     &    0.288  &     2.5    &       6.7    &       3   &     8.8$^c$ &   \\
 J151640.22+001501.8     &   0.0526  &    3.42    &       6.7    &    2.33   &     8.8$^c$ &   \\
 J155522.04+281323.1     &    0.149  &    2.06    &       6.9    &    2.43   &     8.8$^c$ &   \\
 J163323.58+471858.9     &    0.116  &    2.05    &       6.4    &    0.92   &       --    &  Type 2 FWHM(H$\alpha$)=990   \\
 J164126.91+432121.6     &    0.221  &    2.19    &         7    &    1.09   &       --    &  Type 2   \\
 J164442.53+261913.2     &    0.144  &     2.5    &       6.6    &    0.62   &     7.6$^d$ &     \\
 J211852.96-073227.5     &     0.26  &    2.58    &       7.1    &    1.48   &     7.69$^a$ &    \\
 J215226.03-081024.9     &   0.0347  &    2.18    &       6.5    &    2.37   &     8.5$^c$ &   \\
\hline
\label{tabnewmasses}
\end{tabular}
\\
  $^a$  Using   H$\beta$  from  \citet{shen10}  and
  applying the correction for radiation pressure. \\
  $^b$  BH mass estimated  using the
  K-band magnitude measured from 2MASS images (this work) and the \citet{marconi03}
  formula. \\
$^c$  BH mass estimated  using the
  K-band magnitude from 2MASS  catalog (as taken from NED) and the \citet{marconi03}
  formula. \\
  $^d$ Radiation pressure corrected, using FWHM of H$\alpha$ from
  \citet{rafter09}.  
\end{minipage}
\end{table*}

Firstly, we search the recent literature for black hole mass estimates
based on different indicators. We find five objects in common with the
sample analyzed  by \citet{shen10},  which perform careful  line width
measurement  after continuum  subtraction.   We use  the  FWHM of  the
H$\beta$ emission  line and  the continuum luminosity  L$_{5100}$ from
that  work,   and  we   estimate  the  BH   mass  using   the  updated
\citet{marconi08}  formula,  which includes  the  effect of  radiation
pressure on the BLR.  The  BH mass estimates obtained with this method
(see  Table~\ref{tabnewmasses}) are  significantly  larger than  those
estimated  by  \citet{rafter09}.   The  use  of  H$\beta$  instead  of
H$\alpha$  also ensures higher  accuracy, since  the H$\beta$  line is
relatively  isolated, and  its relation  with  the BH  mass is  better
calibrated.   That  is especially  true  for  AGNs  of relatively  low
luminosity, where we do not  anticipate a strong contribution from the
FeII lines in the H$\beta$ spectral region.

We could  not find other BH  mass estimates in the  literature for the
remaining 31 objects.   However, the large majority of  these AGNs are
associated  with   low  redshift  bright   early-type  galaxies.   For
early-type  galaxies, which  typically show  red colors  \citep[$u-r >
2.22$,][]{strateva01}  compatible with an  old stellar  population, we
can estimate the mass of the black hole using the correlation with the
K-band  near IR  luminosity  \citep{marconi03}, as  obtained from  the
2MASS catalog (NED).
We can apply such a method to eighteen objects in the sample.  For two
out of these eighteen objects  the K$_s$-band magnitude from the 2MASS
is not available from NED, therefore we download the fits images and we
measure the magnitude from the  images.  The BH masses we derive using
the  correlation with  the IR  magnitude  are all  larger than  $10^8$
M$_{\odot}$ (see  Table~\ref{tabnewmasses}).  One further  object with
$u-r > 2.22$ (namely  SDSS J090307.84+021152.2) is a Type~2 QSO/ULIRG,
therefore its K-band flux might be contaminated by emission related to
the hot dust  surrounding the AGN.  We checked  the HST/WFPC2 image of
that object and at  8100\AA\ the galaxy morphology appears irregular,
with possible presence  of dust and star forming  clumps.  If the
2MASS  flux  is  used,  its   estimated  BH  mass  is  $2\times  10^9$
M$_{\odot}$,  but  we  believe  that  such  a  value  is  most  likely
unreliable.

We still have to check the 12  objects for which the BH mass cannot be
reliably estimated  using the K-band  magnitude because of  their blue
colors, and  for which  no other  BH mass estimates  are found  in the
literature.

For  J142237.91+044848.5  and  J163323.58+471858.9  the  FWHM  of  the
H$\alpha$ line is  smaller than 1000 km s$^{-1}$,  which is often used
as  the threshold  values  between Type~1  and  Type~2 AGN  \citep[see
e.g.][]{antonucci93,urrypad95}. In  those objects, the mass  of the BH
cannot be reliably estimated from those measurements, unless the width
of  the forbidden  lines is  significantly  smaller than  that of  the
permitted  lines.  From  visual  inspection of  the  SDSS spectra  (and
``quick and dirty'' line fitting) that does not appear to be the case.
The detected  emission lines are  most likely produced in  large scale
regions not under the direct influence of the black hole gravitational
field.

J075444.08+354712.8 and  J115409.27+023815.0 are ``bona  fide'' Type~1
AGNs (i.e.   the measured FWHM  of H$\alpha$ is $>  2000$km s$^{-1}$).
The former is a bright quasars  and the latter is either classified as
a  Sy~1, or as  an FSRQ.   The BH  masses can  be estimated  using the
Marconi et al.   relation.  Assuming FWHM(H$\alpha$) = FWHM(H$\beta$),
we derive  are $\sim 6\times  10^7$ M$_\odot$ and $\sim  5\times 10^7$
M$_\odot$ for the two objects, respectively. 

Careful inspection of the SDSS spectra of the other three objects with
quoted    FWHM(H$\alpha$)   $>2000$    km   s$^{-1}$    reveals   that
J140638.22+010254.6  has both permitted  and forbidden  emission lines
with prominent blue wings, possibly  indicative of an outflow which is
most likely not produced  within the BLR.  For J094525.90+352103.6 and
J103915.69-003916.9  the  classification  as  broad  line  objects  is
extremely  uncertain,  and  they   are  in  fact  both  classified  as
``galaxy'' in the SDSS.

The  remaining five  galaxies  have FWHM(H$\alpha$)  between 1000  and
2000km s$^{-1}$.   For J164442.53+261913.2, after  taking into account
of  the  effects of  radiation  pressure,  the  estimated BH  mass  is
$10^{7.6}$   M$_{\odot}$.   J092936.73+571149.8,  J135646.10+102609.0,
J144341.53+383521.8, and  J164126.91+432121.6 are Type~2  objects.  In
fact, the  SDSS spectra show that  the [OIII]5007 line is  the same as
(or even slightly broader than) the permitted lines.  This most likely
implies that the  line emission region lies outside  the BLR, and thus
out of the sphere of influence of the black hole.

Summarizing, a  careful inspection of  the large sample of  low-z AGNs
from the  SDSS shows  that there  is no clear  evidence for  {\it bona
fide}   radio  loud   objects  associated   with  black   hole  masses
significantly  smaller than  $\sim  10^8$ M$_\odot$.   The smallest  BH
associated   with    a   radio-loud    object   we   find    is   SDSS
J115409.27+023815.0, for  which the estimated BH  mass is $\sim 5\times
10^7$ M$_\odot$. In  other words, even the lowest  BH mass we estimate
is still compatible with $10^8$ M$_\odot$, within the typical error.

\subsection{On the impact of the radiation pressure correction}
\label{onmarconi}

In the  previous paragraphs  we have described  the methods we  use to
estimate the BH  masses for various objects in  the different samples.
In   doing  so,   we  used   the   most  updated   formulae  and   the
\citet{marconi08}  correction  to take  into  account  the effects  of
radiation pressure  onto the BLR.   Although we strongly  believe that
that  is the  correct approach,  we must  point out  that  the overall
results of this  paper are unaltered if we  neglect radiation pressure
effects.

In fact,  using \citet{vestpeterson06} for the SDSS  objects for which
we used  the \citet{marconi08} formula  (the ones marked with  "a" and
"d" in Table 1), we still obtain that the lowest BH mass is $10^{7.6}$
M$_{\odot}$ (for J115409.27+023815.0).   This is because those sources
are  all  low luminosity  AGNs,  therefore  the  effects of  radiation
pressure are small.  To be precise, the Marconi formula, in that case,
returns an even smaller value of the BH mass than the one obtained with
the Vestergaard formula, which  neglects radiation pressure effects at
all luminosities.

Among the  samples of AGN considered  in this paper,  the only objects
that would be inconsistent (assuming a factor of $\sim 3$ error on the
BH mass  estimate) with  the "limit" at  $10^8$ solar masses  are four
QSOs in the  \citet{gu} sample.  However, for those  four objects, the
classification as  Type~1 AGNs is  extremely uncertain.  Three  out of
those four are  in fact classified as Type~2  (NED) and one (1045-188)
is  an  FSRQ.  Going  back  to the  original  paper  that reports  the
spectrum  of that  object \citep{stickel93}  and the  values  used for
deriving its BH mass, we see that the FWHM of H$\beta$ is quoted to be
smaller than that of the  [OIII]5007 line.  Furthermore, a note states
that the  H$\beta$ line is  blended with some  athmospheric absorption
features.  We  conclude that even without  using the \citet{marconi08}
correction, there is no clear  evidence for radio loud AGNs associated
with masses smaller than $\sim 10^8$ M$_\odot$.  

\section{Discussion and conclusions}
\label{discussion}

Using the most updated black  hole mass estimators, we have shown that
there is  no evidence  for a population  of radio-loud  AGN associated
with  supermassive  black holes  of  M  $<  \sim 10^8$  M$_\odot$,  in
agreement   with  previous   work  performed   on   different  samples
\citep[e.g.][]{laor00,dunlop03,papllagn,best05,baldicapetti10}.
Building on this  finding, we propose that the  RQ/RL dichotomy can be
explained by  a modification of the  spin paradigm in  which the radio
loudness of an AGN is determined not only by the spin, but also by the
mass of  the SMBH.  RL AGN are  only those with BH  masses larger than
$\sim 10^8$ M$_\odot$.   Clearly, it is still possible  that the value
of $10^8$ M$_\odot$ does not correspond to any specific threshold, and
it could  just represent a typical  value below which  the probably of
having a radio loud source becomes increasingly small.

While a discussion on the physics  of the jet production is beyond the
scope of  this paper,  in the following  we will present  evidences in
support of  our conjecture, and  we will discuss the  consequences for
the relations between BHs and their host galaxies.

First of all, it is clear that  the mass of the black hole must play a
role, at all AGN luminosities, because the lack of radio loud AGN with
small BH masses is apparent and it is not due to any trivial selection
bias.  Second of all, it is  important to note that for high BH masses
both RL  and RQ AGNs  exist.  Thus, not  surprisingly, the BH  mass is
clearly not  the only physical  parameter involved in  determining the
level  of  radio  loudness  of  each object.   The  so  called  ``spin
paradigm''  \citep{blandfordsaasfee,wilsoncolbert95}  has  often  been
used to explain the RQ/RL  dichotomy.  In brief, assuming that the jet
power  is   related  to   the  BH  spin   ($J\sim  (a/M_{BH})^\alpha$,
\citealt{blandfordznajek77,tchekhovskoy10}),  RL AGNs are  explained as
objects  powered by  rapidly spinning  black  holes.  \citet{sikora07}
have recently proposed a modified  version of the spin paradigm, which
includes two additional elements: i)  the spin of the BH in elliptical
galaxies  can reach  higher values  with respect  to that  of spirals,
because  of their  different merger  history; ii)  {\it only}  at high
accretion rates,  intermittency of jets  collimation causes an  AGN to
switch between  radio-loud and  radio-quiet states.  According  to the
scenario  proposed  by  those  authors,  powerful RQ  QSOs  hosted  by
ellipticals possess rapidly  spinning black holes as the  RL QSOs, but
they are in a state in which  the jet is not collimated. In that case,
the host  galaxies of  RL and RQ  QSO should be  indistinguishable, as
well  as their  large-scale  environment. However,  there is  mounting
evidence that  the RL QSOs  live in significantly  richer environments
than RQ QSOs \citep[e.g][]{shen09,donoso09}.

Here we  propose that the  RQ/RL dichotomy  can be  explained by  a further
modification of the spin paradigm, based on our finding that the
mass  of the  black hole   plays a  role. Investigating  the
physical reasons for that is beyond the aim of this paper. However, we
note that the  BH mass is in fact intimately  related to the accretion
and ejection region  around the BH itself, as it  sets both the radius
of the  innermost stable orbit and the  critical Eddington luminosity,
as originally pointed out by \citet{blandfordsaasfee}.

In  our  conjecture, the  spin  of  the black  hole  plays  a role  in
determining the  radio loudness {\it  only} if the  mass of the  BH is
$\sim 10^8$  M$_\odot$ or higher. For  a smaller BH mass,  the spin of
the BH is irrelevant, as  the Seyfert~1 galaxy MCG--6--30--15 and objects
similar to  that appear to  show. MCG--6--30--15 probably  represents the
best    case   of    maximally   rotating    black   hole    in   AGNs
\citep[e.g.][]{iwasawa96}, and  it has  often been used  to contradict
the  ``spin paradigm''. The  object is  a well  known radio  quiet AGN
hosted by  a E/S0 galaxy, and  its estimated black hole  mass is $\sim
6\times 10^6$ M$_\odot$ \citep{mchardy05}, over 1 dex smaller than any
radio  loud AGN.   We speculate  that  the reason  why MCG--6--30--15  is
radio-quiet ultimately resides in the  fact that its back hole mass is
not  large enough  to  produce a  radio  loud AGN  (i.e.   to power  a
powerful relativistic jet), even if its BH is maximally rotating.  The
same  argument   can  be  applied  to  other   Seyferts  with  similar
properties,   e.g.   1H~0707-495   \citep{fabian09},  which   is  also
associated with a BH mass  of order $10^6$ M$_\odot$.  In other words,
the   objects  that   are  often   used  as   ``exceptions''   to  the
spin-paradigm, are simply indicating that for an AGN to be radio loud,
not only the BH has to be spinning, but also its mass must be close to
or above $10^8$ M$_\odot$.  Therefore, our conjecture accounts for the
fact that radio  quiet AGNs can host black holes  of all masses, while
RL AGNs cannot.

At high BH masses (above $\sim 10^8$ M$_\odot$) the spin regulates the
radio loudness, as in the  original version of the spin paradigm, with
the RQ QSO  having slowly rotating (or non  rotating) black holes, and
the  RL  QSOs  having  rapidly  rotating  black  holes.   Whether  the
transition occurs ``sharply'' at some  particular value of the BH mass
or it  is instead  a rather  smooth transition is  not clear  from the
data.  But  the lack of  RL AGN below  $\sim 10^8$ M$_\odot$  seems to
indicate that  the existence of  a sharp ``threshold''  value for
the BH mass cannot be ruled out.

Another interesting  aspect is to  explore the connection  between the
radio loudness  of the  active nucleus and  the structure,  origin and
evolution of  the host galaxy.  In  low redshift AGNs there  is a well
established  dichotomy  in the  properties  of  the radial  brightness
profiles  or  AGN hosts:  RL  nuclei  are  invariably associated  with
``core'' galaxies,  while RQ nuclei reside in  ``power-law'' or spiral
galaxies,     which    in    turn     have    a     power-law    bulge
\citep{sandrobarbara06,sandrobarbara07}.  \citet{deruiter05} have also
pointed out  that radio galaxies  are invariably associated  with core
galaxies.  This  is indicative  of a profound  link between  the RQ/RL
dichotomy and the  history of the host galaxy,  as originally noted by
\citet{sandrobarbara06}.   It has  been  argued \citep[e.g.][and  ref.
therein]{faber97,merritt06, kormendy09} that core galaxies most likely
originate in major  dry mergers.  The binary black  hole formed during
the merger ejects stars away  from the central regions, thus producing
the  observed stellar  light deficit.   On the  other  hand, power-law
galaxies may originate  in wet mergers, which triggers  a starburst to
create the ``extra-light'' at the center of the galaxy.

Somehow,  the  process  of  formation  of  core  galaxies  has  to  be
associated   with  rapidly   spinning  black   holes  of   high  mass.
\citet{hughesblandford03} have  shown that mergers of  two black holes
of  different mass  may originate  a spinning  BH only  for particular
values  of ``plunge  inclination''  and  only if  the  original BH  is
already spinning  \citep[see also][]{bertivolonteri08}. Otherwise, the
BH is spun-down.  Therefore, rapidly  spinning BHs are unlikely to have
suffered a recent  minor merger. On the other  hand, two merging black
holes  of similar  mass may  produce  a rapidly  spinning black  hole.
Another possible mechanism to  ``spin-up'' a SMBH is through accretion
of matter onto the BH \citep{volonteri07}, although that probably only
leads to  moderate spin  values \citep{king08}. Therefore,  major {\it
dry}  mergers   seem  to  perfectly  fit  the   requirements  both  to
``spin-up'' the black hole, and to originate ``core'' galaxies.

On the  other hand, a gas-rich  ({\it wet}) merger may  provide gas to
fuel  a central  starburst,  which in  turn  produces the  extra-light
observed in  power-law galaxies \citep[e.g.][]{kormendy09}.   Major wet
mergers  are more  likely to  involve galaxies  with  relatively small
bulge mass, and  thus hosting small mass black  holes. Therefore, even
if the resulting spin of the black hole may be significant (because of
either gas accretion or merger of  two BHs of similar mass), the total
BH mass is still not sufficient to power a radio-loud AGN.  That might
be the case for S0 galaxies such as MCG--6--30--15.

Recently, \citet{dotti10} have proposed  a similar scenario. One major
difference  resides   in  the   fact  that  these   authors  associate
radio-loudness  with counter-rotating  accretion  on rapidly  spinning
black  holes resulting  from major  dry mergers.  Counter  rotation is
necessary to increase the  efficiency of jet production.  However, one
possible complication is that the existence of a significant number of
radio-quiet AGN  associated with core  galaxies is expected in  such a
scenario   (those   with   co-rotating   accretion).    Instead,   the
observations show that,  at least at low redshifts,  those objects are
missing.

Summarizing,  our  conjecture  implies  that the  RQ/RL  dichotomy  is
strictly linked  to the history  of the host galaxy,  independently of
the  accretion  rate.  In  order  to  produce  a radio-loud  AGN,  two
conditions have to  be satisfied: 1) $M_{BH} >  \sim 10^8 M_\odot$; 2)
the spin of the BH must be significant. Major {\it dry} mergers of two 
galaxies, whose
black  holes have  masses close  to or  above $10^8$  M$_\odot$,  lead to
radio-loud  AGNs.  Smaller  mass BHs  cannot produce  a  powerful jet,
therefore the  actual BH  spin is unimportant  (or less  important) in
those objects, as far as the radio loudness is concerned.

An obvious criticism is that some stellar mass black holes are capable
of producing rather powerful radio jets. Although these objects can be
bright  in  the  radio,  it  is important  to  estimate  their  actual
radio-loudness.   The  problem in  this  case  is  that their  optical
emission  is generally not  observed, for  various reasons\footnote{In
most  cases,  the  optical  emssion  of  these  galactic  binaries  is
difficult  to  study  and  may  be  produced  by  different  competing
mechanisms, most  importantly by the  companion star.  In  addition to
that,  in the  case of  the Galactic  superluminal  GRS1915+105, heavy
absorption (A$_V \sim 44$mag) completely hides the optical source.}.
However, we  can estimate the radio loudness  using the radio-to-X-ray
luminosity ratio, as proposed by \citet{terashima03}.  For example, in
the  case of  GRS1915+105, $\log  R_x =  \log (\nu  L_{5GHz} /  L_X) =
-6.6$, while the  typical value for radio loud AGNs  is $R_x \sim -1$,
and  even the  radio  quiet AGNs  have  $R_x \sim  -3$ \citep[see  the
compilation  of  data in  e.g.][]{merloni03}.  These  objects are  not
``radio-loud'',  even  if they  produce  a  rather  bright radio  jet.
Therefore, besides all of the differences that might exist between the
properties of  the environment in  the vicinity of  supermassive black
holes and that of stellar mass black holes, which might as well play a
role  in the  physical properties  of the  outflows, the  stellar mass
black holes  are significantly  less radio loud  than all  AGNs.  This
confirms our findings, i.e. small mass BH are not capable of producing
true radio loud objects.


Differently  from  \citet{sikora07}, and  building  on  the results  of
\citet{sandrobarbara06},  in our proposed  scenario the  properties of
the host  galaxy are strictly connected  to the radio  loudness of the
nuclei,  via  their  formation  and evolution,  independently  of  the
current accretion  rate level.  In this way,  no ``intermittency'' and
switching between RQ and RL ``states'' for high accretion rate AGNs is
needed,  as it  seem to  contradict  the result  that RQ  and RL  QSOs
inhabit different environments, as  discussed above.  Therefore, in our
picture, there is no difference  between low and high luminosity AGNs,
and low and high accretion rate (or Eddington ratio).

Such a scenario implies a straightforward prediction: differently from
\citet{sikora07},  we expect  powerful radio-quiet  QSOs to  reside in
power-law galaxies, while their RL counterparts would be associated to
core  galaxies,  exactly  as  observed  at  low  redshifts.   Clearly,
confirming such a prediction is no  easy task, due to the high angular
resolution needed to resolve  cores in distant galaxies, especially in
presence of a bright quasar  nucleus.  Obscured QSOs might be a better
choice for  that, under  the assumptions that  type~1 and  type~2 AGNs
belong       to       the       same       ``parent''       population
\citep{antonucci93,urrypad95}.  However,  even in that  case, the core
radius in a z=0.5 object would be only $\sim 10$ mas. Therefore, other
structural parameters of  the host galaxy that are  found to correlate
with the properties  or the innermost structure are  more likely to be
observed   with   current  instruments   and   in   the  near   future
\citep[e.g. the Sersic index,][]{kormendy09}.

Another implication of our proposed  scenario is that broad iron lines
bearing the signature  of the Kerr metric (as  seen in MCG--6--30--15)
should be  present in the  X-rays in radio loud  (lobe-dominated) QSOs
and in broad line radio  galaxies.  These are objects that are thought
to  be seen  at the  right  orientation to  allow direct  view of  the
accretion disc, while the  relativistically beamed jet emission should
be observed at an angle that  is large enough for that not to dominate
the overall  emission.  Therefore, the inner regions  of the accretion
disc should  be visible, unless that  region of the  accretion disc is
``emptied'' because of the presence  of significant outflows, or if it
is in an  ADAF state \citep[see e.g.][]{yuan07,tchekhovskoy10}.  Until
now,  no  Fe lines  of  that kind  have  been  detected in  radio-loud
objects,  to  the best  of  our knowledge.   But  that  might just  be
explained  by the  low sensitivity  of current  instrumentation, which
does not allow a clear detection of those features in distant objects.

An obvious way of falsifying our conjecture is the observation of such
relativistically  broadened   Fe  lines  in  the   X-ray  spectrum  of
radio-quiet  QSOs associated  with a  large  black hole  mass. Such  a
result  would disproof the  hypothesis that  a rapidly  spinning black
hole with a mass larger than $10^8$ M$_\odot$ is sufficient to produce
a radio loud AGN.

\section*{Acknowledgments}

We are  grateful to Mario Livio, Marta  Volonteri, Alessandro Capetti,
Colin Norman  and Gabriele  Ghisellini for insightful  discussions. We
also thank  the referee for  her/his valuable comments that  helped to
improve  the paper.   This  research  has made  use  of the  NASA/IPAC
Extragalactic Database  (NED) which is operated by  the Jet Propulsion
Laboratory,  California Institute of  Technology, under  contract with
the National Aeronautics and Space Administration.


\bsp

\label{lastpage}

\end{document}